\documentclass[12pt]{article}

\usepackage{scicite}

\usepackage{times}
\usepackage[utf8]{inputenc}
\usepackage[T1]{fontenc}
\usepackage{comment}
\usepackage{subcaption}
\usepackage{subcaption}
\usepackage{graphicx}
\usepackage{graphics}
\usepackage{lineno}

\newenvironment{sciabstract}{%
\begin{quote} \bf}
{\end{quote}}

\title{A Deep-Learning Usability Expansion Model of Ocean Observations}

\author
{Ali Muhamed Ali,$^{1\ast}$ Hanqi Zhuang,$^{1}$ Yu Huang,$^{1}$ \\
Ali K. Ibrahim,$^{2}$ Ali Salem Altaher,$^{1}$ Laurent Ch\'erubin$^{2}$\\
\\
\normalsize{$^{1}$Electrical Engineering and Computer Science Department, Florida Atlantic University,}\\
\normalsize{ Boca Raton, FL, USA,}\\
\normalsize{$^{2}$Harbor Branch Oceanographic Institute, Florida Atlantic University, FL, USA}\\
\\
\normalsize{$^\ast$Ali Muhamed Ali; E-mail:  amuhamedali2014@fau.edu.}
}

\date{}

\begin{document} 


\baselineskip24pt


\maketitle


\begin{sciabstract}
  Today's ocean numerical prediction skills depend on the availability of in-situ and remote ocean observations at the time of the predictions only. Because observations are scarce and discontinuous in time and space, numerical models are often unable to accurately model and predict real ocean dynamics, leading to a lack of fulfillment of a range of services that require reliable predictions at various temporal and spatial scales. The process of constraining free numerical models with observations is known as data assimilation. The primary objective is to minimize the misfit of model states with the observations while respecting the rules of physics. The caveat of this approach is that measurements are used only once, at the time of the prediction. The information contained in the history of the measurements and its role in the determinism of the prediction is, therefore, not accounted for. Consequently, historical measurement cannot be used in real-time forecasting systems. The research presented in this paper provides a novel approach rooted in artificial intelligence to expand the usability of observations made before the time of the prediction. Our approach is based on the re-purpose of an existing deep learning model, called U-Net, designed specifically for image segmentation analysis in the biomedical field. U-Net is used here to create a Transform Model that retains the temporal and spatial evolution of the differences between model and observations to produce a correction in the form of regression weights that evolves spatially and temporally with the model both forward and backward in time, beyond the observation period. Using virtual observations, we show that the usability of the observation can be extended up to a one year prior or post observations.
\end{sciabstract}


\section*{Introduction}
Today's ocean numerical prediction skills relies on the availability of in-situ and remote ocean observations at the time of the predictions only. Because observations are scarce and discontinuous in time and space, numerical models are often unable to accurately model and predict real ocean dynamics, leading to a lack of fulfillment of a range of services that require reliable predictions at various temporal and spatial scales. To this end, it is crucial to have sustained and reliable observations of the parts of the ocean where the predictions are made. The process of constraining free numerical models with observations is known as data assimilation. The primary objective is to minimize the misfit of model states with the observations while respecting the rules of physics. Observations assimilated in ocean forecasting systems now include altimetry, ocean colour, surface velocities, sea ice and data from emerging platforms such as ocean gliders. Many systems now employ multi-model approaches or ensemble modeling techniques. A variety of data assimilation are used in operational oceanography and a detailed review of the data assimilation methods can be found in \cite{Moore2019}. The most common data assimilation method closely parallels the procedures employed in numerical weather prediction \cite{kalnay2003} in which ocean state estimates are computed sequentially through time, and the resulting estimates updated when sufficient new observations become available. The caveat of this approach is that measurements are used only once, at the time of the prediction. The information contained in the history of the measurements and its role in the determinism of the prediction is, therefore, not accounted for. Consequently, historical measurement cannot be used in real-time forecasting systems. 
 
 Because oceans play a key role in global issues such as climate change, food security, and human health, effort to create long term repositories with global quality assurance/quality control standards have received significant support over the last decade. These efforts have come to fruition in the Global Ocean Observing System (GOOS) \cite{jager1991}. GOOS has opened a field of opportunity for new collaborations across regions, communities, and technologies facilitating enhanced engagement in the global ocean observing enterprise to benefit all nations. In the United States, these efforts led to the creation of the U.S. Integrated Ocean Observing System (IOOS), which through its eleven Regional Associations implement regional observing systems covering all U.S. coasts and Great Lakes with activities spanning from head of tide to the U.S. exclusive economic zone \cite{Snowden2019}. Through these efforts comprehensive ocean datasets, spanning several decades have become available to the research community and novel technologies rooted in big data science such as Artificial Intelligence (AI) have emerged \cite{Trice2021}. The research presented in this paper provides a novel approach rooted in AI to expand the usability of observations made before the time of the prediction. 
 
To demonstrate our method, we take advantage of a recent and comprehensive set of observations in the Gulf of Mexico (GoM), which has received over many decades, and continues to receive, a lot of attention by oceanographers. The GoM ocean dynamics is controlled by the pulsating Loop Current, which is the most energetic circulation feature of the basin. Predicting the Loop Current evolution is fundamental to almost all aspects of the GoM, including (a) anthropogenic and natural disaster response; (b) the prediction of short-term weather anomalies and, hurricane intensity and trajectories; (c) national security and safety; and (d) ecosystem services \cite{walker2011,reusch2005ecosystem}. The recent Deep Water Horizon oil spill in 2010 has bolstered a large amount of research toward improving our understanding and forecasting of the Loop Current System in hope of mitigating further environmental and ecological damage. Furthermore, because of the reinforcing interaction between hurricanes, the Loop Current and its eddies, long-term prediction of the Loop Current states is becoming more and more relevant \cite{JAIMES2016306},\cite{oey2006loop}. Thus, developing accurate and robust medium-term forecasting models of the Loop Current System and its eddy formation is imperative and relies on the availability of ocean observations. Improved ocean observing systems are expected to reduce the uncertainty of ocean/weather forecasting and to enhance the value of ocean/weather information throughout the Gulf region \cite{Kaiser2004}.
 
The remainder of the paper is organized as follows. In Section 2 \ref{sec:Sustained temporal correction model} we provide an overview of the concept of data usability temporal expansion. Section 3 \ref{sec:Dataset} describes the numerical models fields and the observations used in this study. The usability temporal expansion concept is applied to both a free and data assimilated simulations. Section 4 \ref{sec:Deep Learning} introduces the deep learning model used in this study. Section 5 \ref{sec:Methodology} presents the implementation of the temporal expansion concepts, which results in the transformation of the ocean models' velocity fields beyond the observation period. Discussion and conclusion follow in Section 6.

\section*{Temporal Expansion Model Concept} \label{sec:Sustained temporal correction model}
The rationale behind this concept lies in the fact that ocean observations in the data assimilation process only serve once, at the time of the forecast, when the background field is created. They are no longer used in forecasts past the time of the observation except in three-dimensional variational data assimilation where all data collected within a short time window (days) are assimilated as though collected at a single time \cite{Moore2019}. Considering, observations are no longer relevant after a few days and become history. However, there is a history of the ocean dynamics captured in the observations that is not part of the constraint provided by the data assimilation concept. The dynamical history in the observations can be captured with deep learning models as shown by prediction exercises of the Loop Current dynamics \cite{justin2019}, \cite{wang2021}, \cite{muhamed2021}, \cite{Huang}. Deep learning allows computational models that are composed of multiple processing layers to learn representations of data with multiple levels of abstraction. These methods have dramatically improved the state-of-the-art in speech recognition, visual object recognition, object detection and many other domains such as drug discovery and genomics. Deep learning discovers intricate structure in large data sets by using the back-propagation algorithm to indicate how a machine should change its internal parameters that are used to compute the representation in each layer from the representation in the previous layer \cite{LeCun:2015dn}. 

Ocean numerical models are mathematical approximations of the laws of physics that are used to estimate ocean states given a set of boundary conditions and constraints, such as the atmospheric and the tidal forcing. When not constrained by data assimilation, the simulated states of the "free-run" are likely to be very different from the the real ocean state, although the main features of the ocean dynamics will be captured and resolved. These features include the mesoscale dynamics, the water masses, the eddy kinetic energy, among others, whose variability and magnitude is likely to be dissimilar to the observed ones, although existent in both  natural and virtual systems. Therefore one can assume that both the simulated and real systems are on a parallel track although different in many ways. While Deep Learning has been mostly applied to a single data set for pattern recognition, we show in this study that it can be re-purposed to learn the differences between the simulated and real systems and their evolution over time and space. The acquisition of that knowledge between a given ocean numerical model simulation and a set of observations in the model domain over a time period concurrent to both, enables the prediction of the differences with a Deep Learning model. Therefore, the knowledge of the differences can be converted into a model field observations-based correction tool, called a Transform Model, just like a data assimilation method, with the advantage of predicting the correction of the model field beyond the period of observation. This methods also accounts for the common history of the model fields and measurements in the correction process. We will show that the improvements to the model fields, namely the reduction of errors is significant, despite the fact that no physical constraints were applied to the correction. This correction method evaluation would certainly benefit from a numerical model integration to enforce the model conservation constraint although they might be different from the ones set by the measurements. However, testing of the transformed field in a numerical model is beyond the scope of this study.  

\begin{figure}[ht]
\centering
\includegraphics[width=\linewidth]{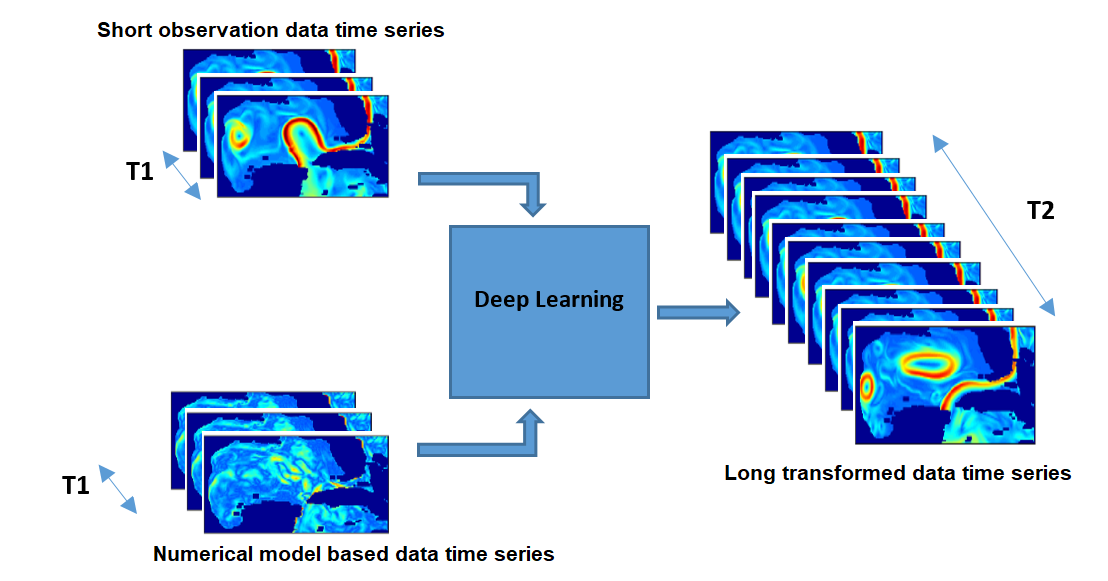}
\caption{Observation usability temporal expansion concept with a Deep Learning approach. Despite the shorter duration of the observations, the overlapping model field can be corrected beyond the observation period.}
\label{fig:Temporal_model}
\end{figure}

The concept of the Transform model is depicted in Figure \ref{fig:Temporal_model}. It is based on the availability of long enough concurrent time series of model field and observations. Both are jointly used in a Deep Learning model that creates a transformation tool of the numerical model field based on the observations. The Transform Model can then be used to correct the numerical model field at any time in the future and in the past, outside of the period of measurements. It thus expands the usability of the observations beyond the measurement period, both before and after.
\section*{Data sets} \label{sec:Dataset}
For the demonstration of our Transform Model we used three data sets that overlap in time, composed of two numerical models current velocity vector fields and one set of in-situ current measurements from an observing array in the GoM. Model's inherent dynamics may influence how effective the Transform Model, which will be shown in Section 5 

The first numerical model data set was obtained from the HYbrid Coordinate Ocean Model (HYCOM) Consortium. The data was generated by the HYCOM $+$ NCODA Gulf of Mexico $1/25^{\circ}$ Reanalysis (GoMu $0.04$/expt\_$50.1$), which spans year 1993 to 2012 and has a horizontal resolution of $4.4$km. HYCOM $+$ NCODA is a data-assimilative hybrid isopycnal-sigma-pressure (generalized) coordinate ocean model. The system uses the Navy Coupled Ocean Data Assimilation (NCODA) system \cite{Cummings2005}, \cite{Cummings2013} for data assimilation. NCODA uses the model forecast as a first guess in a three-dimensional variational scheme and assimilates available satellite altimeter observations (along track obtained via the NAVOCEANO Altimeter Data Fusion Center), satellite and in-situ Sea Surface Temperature (SST) as well as available in-situ vertical temperature and salinity profiles from XBTs, ARGO floats and moored buoys. The Modular Data Assimilation System (MODAS) is used for downward projection of surface information \cite{Fox2002}. The experiment includes tidal forcing. As noted here, no current vector flow information is used in the data assimilation process. 

The second model's current vector fields were obtained from a free-running Massachusetts Institute of Technology (MIT) generalized circulation model simulation of the GoM circulation (MITgcm-GoM), a z-level model. The MITgcm–GoM model was originally developed for state estimation and prediction of the upper ocean circulation in the GoM, including Loop Current evolution and eddy shedding. For these purposes, satellite-derived ocean surface observations and subsurface in situ observations were assimilated using a four-dimensional variational (4DVAR) method \cite{gopalakrishnan2013adjoint}. For the data set used in this study, however, no data assimilation is performed. The model uses a telescopic grid with a horizontal resolution of 1/20° × 1/20° in the central GoM which decreases to 1/10° × 1/10° toward the boundaries and western part of the domain. For this study, daily output from the simulation for the four years forced by atmospheric and lateral boundary condition data from 2009 to 2012 are used \cite{Morey2020AssessmentON}. No tidal or atmospheric pressure forcing is applied. 

The in-situ data were obtained from the Dynamics of the Loop Current in U.S. Waters experiment (hereafter Dynloop), a comprehensive observational study of the Loop Current in the eastern GoM \cite{hamilton2016loop}, \cite{donohue2016a}, \cite{DONOHUE2016b}. The observational array consisted of the instrumentation of nine tall-moorings and seven short-moorings, and an array of 25 pressure-equipped inverted echo sounders (PIES). The array system measured the water column velocity for 2.5 years, beginning in March 2009. Each tall-mooring included water velocity measurements made from an upward facing 75 kHz ADCP deployed near a depth of 450 m, as well as point current meters at five depths from around 600-3000 m, providing high resolution water velocity data between about 60-440 m depths and much lower resolution water velocity data below this. Additionally, each short-mooring included a single point current sensor located 100 m above the sea floor, providing near bottom water velocity data. The PIES array provided direct measurement of pressure and acoustic round trip travel times from the sea floor to the sea surface, which were used to create vertical profiles of density, salinity, and temperature. These pressure records combined with estimated horizontal density gradients were used to calculate geostrophic water velocities. This array was located to cover both east and west sides of the Loop Current between the West Florida Slope and the Mississippi Fan, and was also centered over the zone where Loop Current eddies typically separate from the Loop Current.  The horizontal separation between moorings was around $50-80$ km and between the PIES sensors was around $40-50$ km. These recorded data were used to construct the measurement-based water velocity matrix used in this study. More details on the creation of the velocity matrix can be found in \cite{muhamed2021}. The horizontal resolution varies between 30 and 50 km and only the first 500 m, corresponding to 26 vertical layers, was used in this study. The time resolution for the velocity data was 12 hours, which corresponds to 1810 data frames for each u and v velocity component. The final matrix dimensions were of 1810 × 26 × 29 × 36 for each component.

\section*{Transform Model} \label{sec:Deep Learning}

 The deep learning network of choice for our Transform Model is the u-shaped Convolution Neural Network (CNN), also known as U-Net \cite{ronneberger2015u}. The typical use of convolutional networks is on classification tasks, where the output to an image is a single class label. However, in many visual tasks, especially in biomedical image processing, the desired output should include localization, i.e., a class label is supposed to be assigned to each pixel. The architecture of this network consists of two parts. The first part is the analysis path called the encoder, which is similar to the CNN architecture widely used for image feature extraction. The second part is the synthesis path called the decoder, where the compact features are expanded to the input dimension. In the image case, the feature set, or code, is decoded back to a (segmented) image of the same dimension as the input image. The name of U-Net is based on the U structure of the network. This type of network has been widely used in medical image segmentation applications \cite{siddique2020u,chacon2018domain}, for land segmentation studies in remote sensing images \cite{li2018deepunet}, for cloud and shadow segmentation \cite{jiao2020refined}, for modeling and prediction of coastal weather events in the Netherlands \cite{fernandez2020deep}, and for coastal wetland classification \cite{dang2020coastal}, among many other applications.

\begin{figure}[ht]
\centering
\includegraphics[trim={0cm  3cm 3cm 3cm},clip,width=\linewidth]{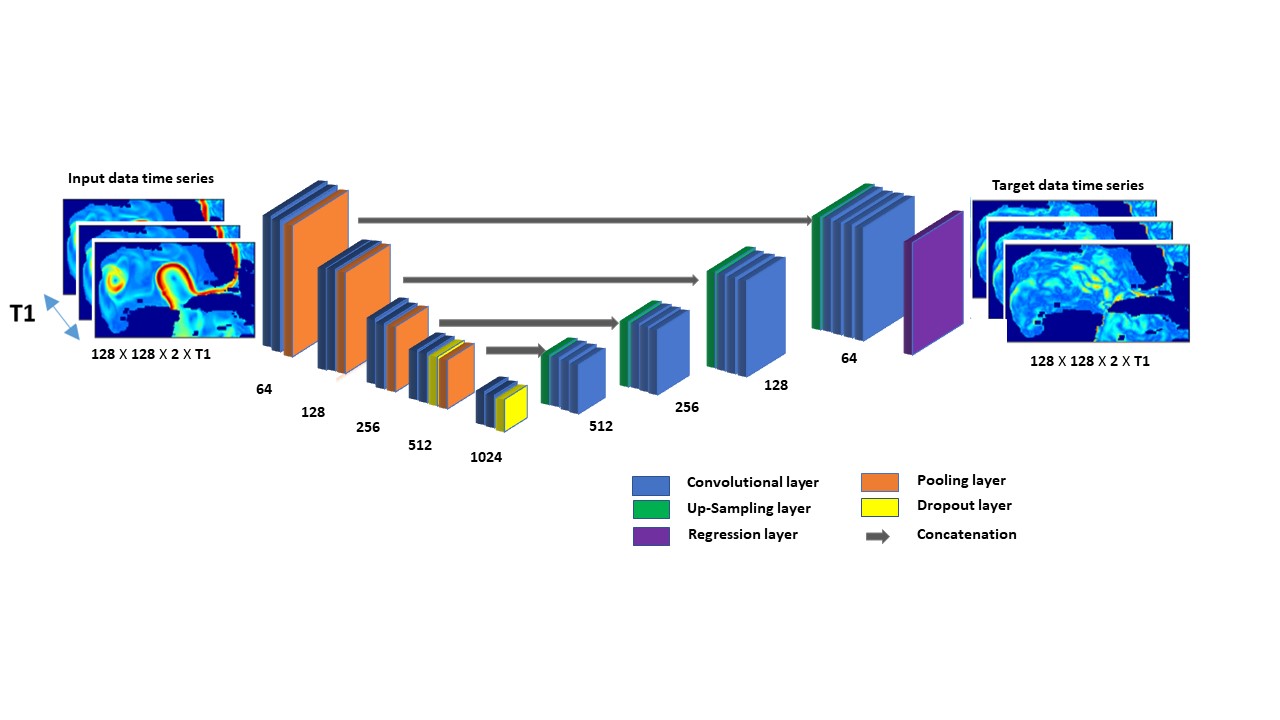}
\caption{The STU-Net structure used in this study for temporal data usability expansion.}
\label{fig:lgraph}
\end{figure}

For our application, the U-Net is slightly modified at the output layers to make it suitable for a regression problem instead of the original structure that was designed for image segmentation. The problem can be seen as a time series transformation through a regression analysis. Because the U-Net will learn both the spatial and temporal features of the training data, it is called hereafter the Spatial Temporal U-Net (STU-Net). Hence, the last layers after the convolution layers have been replaced by a regression layer, instead of a softmax and a segmentation layer in the original structure. The final model structure consists of four stages of encoders as shown in Figure \ref{fig:lgraph}. The input is made of a four-dimensional input layer, which is organized as $x,y,u/v,t$. Here $x,y$ are the coordinates of the sample location, $u/v$ are the zonal/meridional velocities respectively, and $t$ denotes the time of the measurements.

\begin{figure}[ht]
\centering
\includegraphics[trim={0 0cm 0 0cm},clip,width=\linewidth]{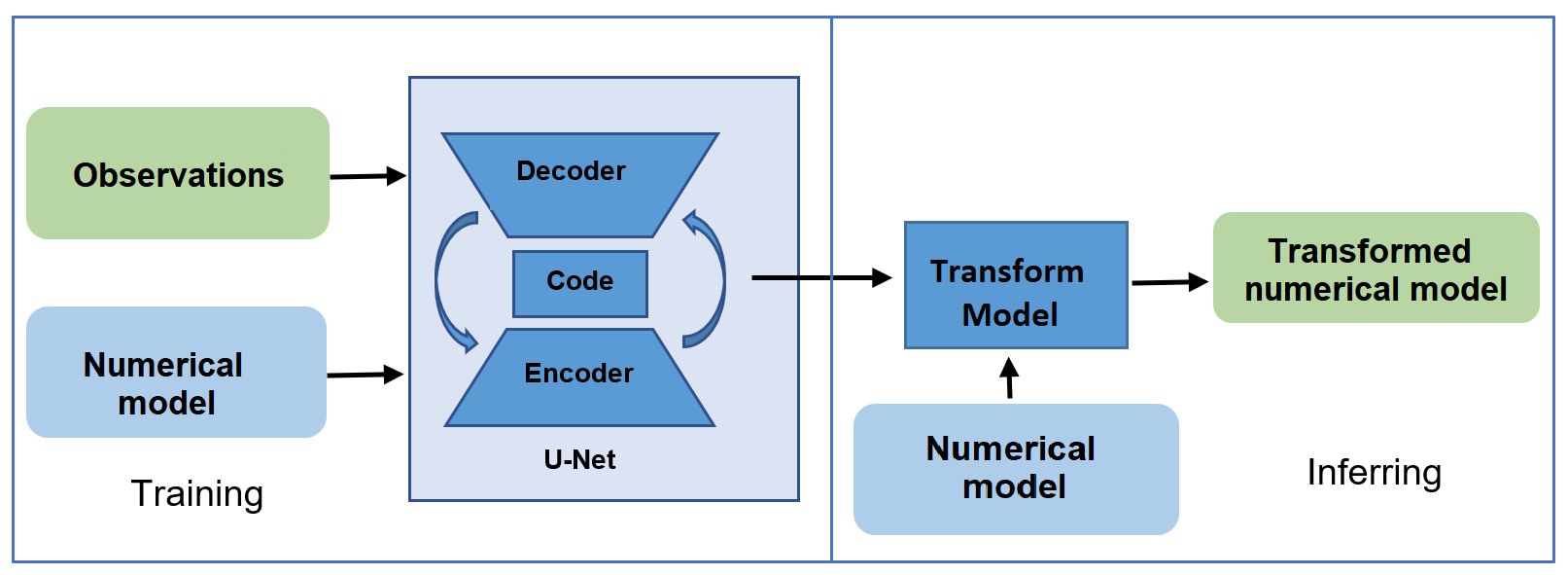}
\caption{Implementation of the proposed Transform Model using a STU-Net architecture. The output layers of the STU-Net, denoted in the figure by "Transform Model", are used to transform the numerical model fields, even when observations are no longer available. }
\label{fig:DL_model_tempora}
\end{figure}

The implementation of the STU-Net is two-fold (Fig. \ref{fig:DL_model_tempora}):  first is the training phase, where both the numerical model and the observations are fed to the STU-Net. The relationship between the measurements and the numerical model is represented by the neural network structure, including its hyper parameters and weights. The errors between the two fields are used to adjust the weights of the model until the training process reaches a certain number of epochs. With a Mini-batch loss set at 0.1 and Mini-batch Root Mean Squared Error (RMSE) of 0.5, 120 epochs were necessary. The cost function limit used in the training stage was set by the RMSE. At the end of the training stage, the obtained regression layers constitute the Transform Model that will be used to correct the numerical model fields before, during and after observations started in the second step of the STU-Net implementation, which is the inferring phase. To apply the transformation back in time, namely before the observation period, training is conducted backward in time to learn the temporal features in reverse order. For the experiments conducted in this study, the hyperparameters were set to be as follows: $ InitialLearnRate = 5e-4, MiniBatchSize = 4, MaxEpochs = 300, LearnRateDropFactor = 0.1.$. This model was implemented with the Matlab Deep Learning Toolbox.

\section*{Implementation of the Transform Model Concept } \label{sec:Methodology}

 
 \subsection{Experimental strategy and input data formatting} 
 In this section the strategy to evaluate our Transform Model concept is presented. First we will assess the transformation of the HYCOM and MITgcm-GoM velocity fields by the in-situ observations. The training is conducted with the first 855 days and the transformation results are evaluated with the remaining 50 days of measurements. The purpose of this experiment is to assess the transformative potential of limited observation periods and its efficacy on assimilated and non-assimilated numerical model fields, respectively. Second, the long-term transformation capability of the concept is assessed by using HYCOM as virtual observations to transform the MITgcm-GoM velocity fields. In this case, the training took place over a period of three years and the testing period was either the preceding or following year of the training period. 
 
 In order to proceed, both the model and the observations must be interpolated on the same spatial and temporal grids. Therefore the time interval was set to daily, MITgcm-GoM's and the coarsest temporal resolution of the three time series. Because the spatial resolution of the Dynloop data set over the geographical area of the measurements is $29 \times 36$, $64 \times 88$ for HYCOM and $52 \times 70$ for the MITgcm-GoM simulations, all three data sets were interpolated to $64 \times 64$ grid using a bicubic interpolation \cite{keys1981cubic}. Vertically, the same depth layers as in the Dynloop dataset were used  to reinterpolate the other two model fields, which resulted in 26 layers of 20m interval from the surface to 500m (maximum depth used in this study).
 
 For the short-term Transform Model, the period March 2009 - June 2011 was used for training and July-August 2011 for testing. For the long-term transformation experiment, the years 2009-2011 (2012-2010) were used for training and 2012 (2009) for testing in the forward (backward) experiment.


\subsection{Comparison between observed and modeled fields} \label{subsec:The input data and training}
\begin{figure}[ht]
\begin{subfigure}{.5\textwidth}
  \centering
  \includegraphics[trim={0 0cm 0 0cm},clip,width=.95\linewidth]{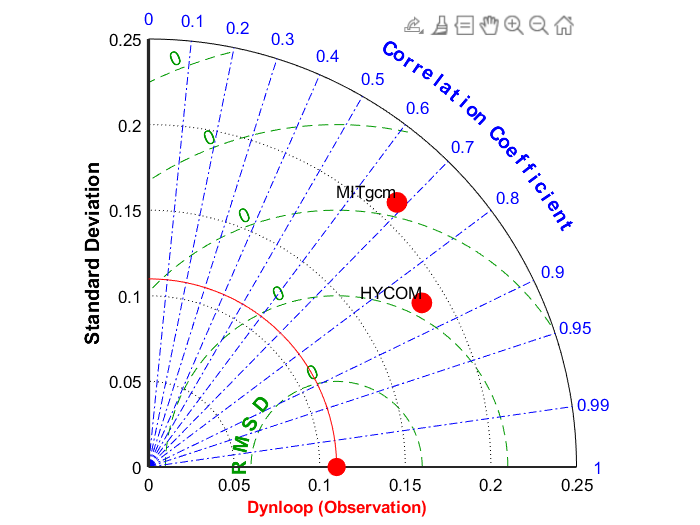}
  \caption{}
  \label{fig:Future_RMSEM}
\end{subfigure}
\begin{subfigure}{.5\textwidth}
  \centering
 \includegraphics[trim={0 0cm 0 0cm},clip,width=.95\linewidth]{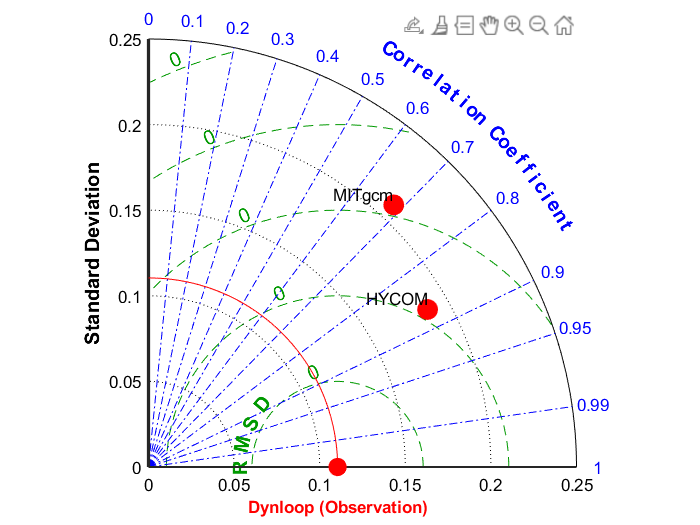}\\   \caption{}
  \label{fig:Past_RMSEM}
\end{subfigure}
\begin{subfigure}{.5\textwidth}
  \centering
 \includegraphics[trim={0 0cm 0 0cm},clip,width=.95\linewidth]{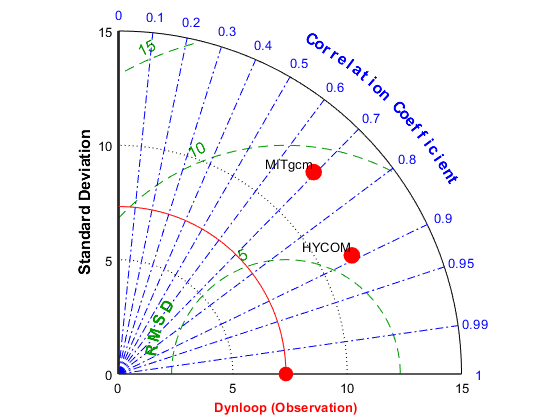}\\ 
  \caption{}
  \label{fig:Past_RMSEM}
\end{subfigure}
\begin{subfigure}{.5\textwidth}
  \centering
 \includegraphics[trim={0 0cm 0 0cm},clip,width=.95\linewidth]{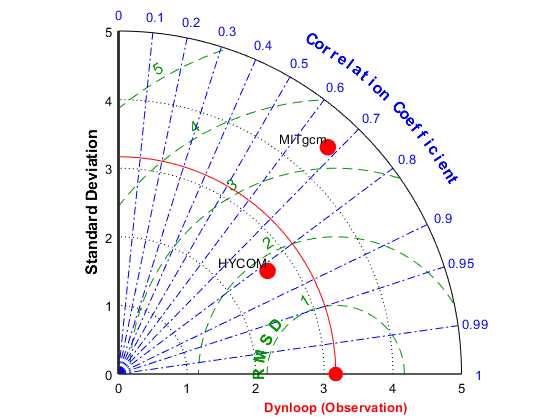}\\ 
  \caption{}
  \label{fig:Past_RMSEM}
\end{subfigure}
\caption{Taylor diagrams of the spatially averaged current velocity magnitude for HYCOM and MITgcm-GoM and Dynloop (observations) at depths of (a) 0m, (b) 20m, (c) 100m, and (d) 500m.}
\label{fig:Taylor}
\end{figure}

To evaluate how different the three concurrent field times series are, we calculated the Taylor diagrams \cite{jolif} of the spatially averaged velocity magnitude for each of them at four depths, 0, 20, 100, and 500m respectively (Figure \ref{fig:Taylor}). Because the MITgcm-GoM is a free-run, the model velocity exhibits at all depths the lowest correlation coefficient (CC), largest RMSE and standard deviation. While the assimilated HYCOM model exhibits a lower standard deviation than the MITgcm-GoM, the RMSE between the two models is not significantly different at 0, 20 and 100m, except at 500m, where HYCOM underestimates it (Figure \ref{fig:Taylor}d). A large RMSE signifies that the amplitude of the variation of the velocity components is overestimated. HYCOM's CC is also lower at 500m than at shallower depth, suggesting that the data correction at 500m is less efficient than in shallower waters.


\subsection{Short-term Transform Model evaluation} \label{sec:ST-TM}
For this evaluation, two types of experiments were conducted. In the first one the transformation was applied only to the two-dimensional surface velocity and in the second experiment the transformation was applied to three-dimensional tensor of the velocity field. The efficacy of the transformation was evaluated by calculating the MSE between the transform model solution and the reference, which is the observations in this case, over the last 50 days of the observational period, not used in the training phase. The same calculation was done for the original numerical model velocity field.

\subsubsection{Two-dimensional field transformation} 
\begin{figure}[ht]
\begin{subfigure}{.5\textwidth}
  \centering
  \includegraphics[trim={0 0cm 0 0cm},clip,width=.95\linewidth]{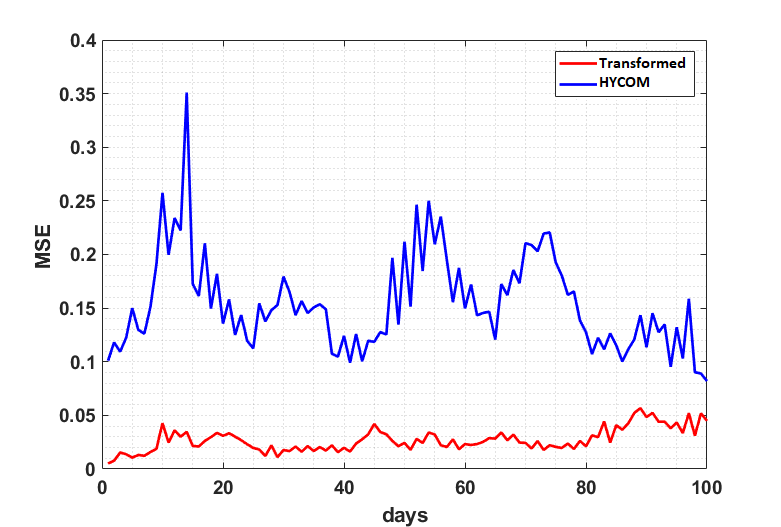}
  \caption{HYCOM  model}
  \label{fig:RMSE_HYCOM}
\end{subfigure}
\begin{subfigure}{.5\textwidth}
  \centering
 \includegraphics[trim={0 0cm 0 0cm},clip,width=.95\linewidth]{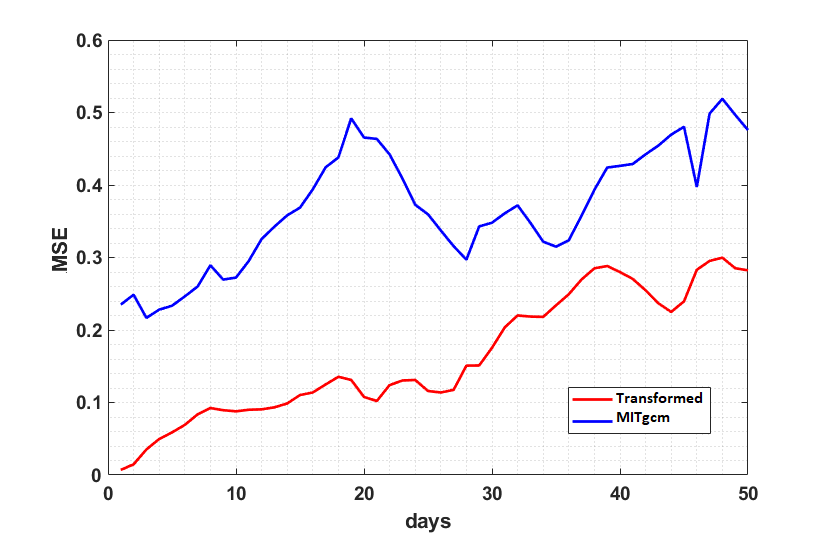}\\ 
  \caption{MITgcm model}
  \label{fig:RMSE_ECCO}
\end{subfigure}
\caption{Surface velocity magnitude Mean Squared Errors (MSEs) between the original and the transformed model for (a) HYCOM model and (b) MITgcm-GoM model}
\label{fig:RMSE_0}
\end{figure}

The results of the MSE for both the original numerical model and the transformed model fields are shown in Figure \ref{fig:RMSE_0} for both HYCOM and MITgcm-GoM.  Figure \ref{fig:RMSE_HYCOM} shows a significant reduction of the HYCOM model velocity MSE, both in the mean and in the variability. The MSE of the transformed HYCOM field remained almost constant throughout the 50 day period, unlike the original HYCOM field. Figure \ref{fig:RMSE_ECCO} shows that the transformation of the MITgcm-GoM fields was as efficient as for the HYCOM fields but the MSE of the transformed fields increased over time, at a rate similar to the one of the original fields. HYCOM's fields' MSE, although quite variable does not exhibit an increasing trend over time like the MITgcm-GoM fields. This difference is most likely due to the fact that data were assimilated in the HYCOM simulation and not in the MITgcm-GoM model. The degree of transformation also called correction gain is shown in Table \ref{tab:correction} and is calculated as follows:

\begin{equation} 
Gain= \frac{ |M_{transformed} -M_{model}|} {M_{model}} * 100
\label{eq:Gain}
\end{equation}
where 

  ($M_{model}= \frac{\sum_{n=1}^{T} { MSE_1(n)}} T$), \quad  
  ($M_{transformed}= \frac{\sum_{n=1}^{T} { MSE_2(n)}} T $) ,\quad   $T$ is the length of the time series, $MSE_1$ is the MSE between the model and the observed data, and the $MSE_2$ is the MSE between the transformed and observed data. A high (low) gain signifies that the transformed model field errors are low (high).

\begin{figure}[h!]
\centering
\includegraphics[trim={0 0cm 0 0cm},clip,width=.9\linewidth]{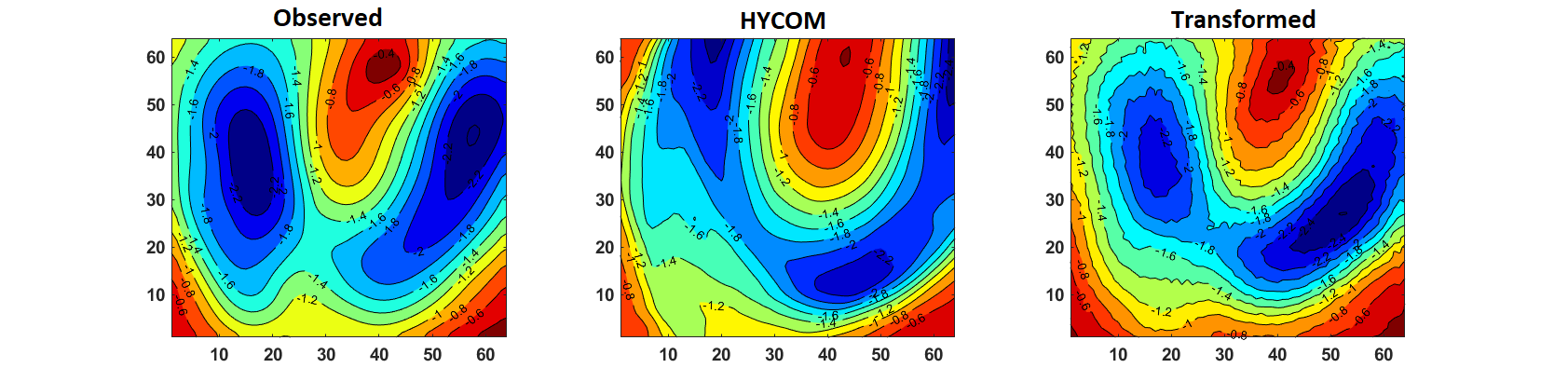}\\
(a) \\
\includegraphics[trim={0 0cm 0 0cm},clip,width=.9\linewidth]{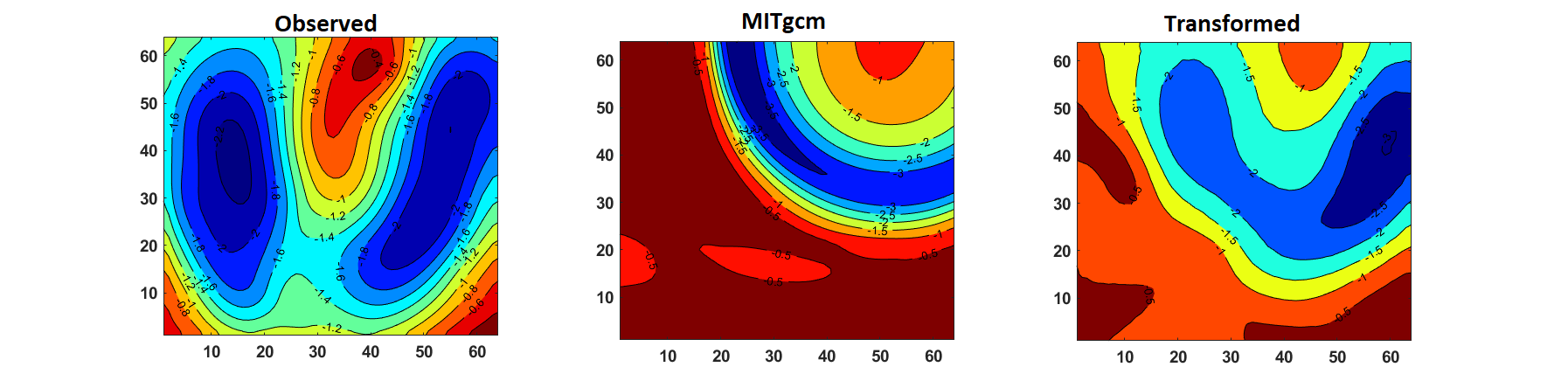}\\
(b) \\
\caption{First Empirical Orthogonal Function (EOF) mode of the surface velocity fields over the 50-day testing period of the Transform Model. From left to right is shown the observed, model original, and transformed field. (a) HYCOM and (b) MITgcm-GoM.}
\label{fig:EOF_0}
\end{figure}

The transformation gain was higher for the HYCOM model than for the MITgcm-GoM field. However, the non-assimilated model transformation exhibited a MSE < 1.5 $cm.s^{-1}$ for up to 30 days, which is the less than the MSE of the assimilated model fields (Figure \ref{fig:RMSE_ECCO}). This result suggests that the correction of the flow field with in-situ observations by the Transform Model can be as effective as if not better than the data assimilation method currently used in the HYCOM operational forecast. This result is significant because the transformation was made outside of the observation period. In terms of the spatial structure of the flow field variability, the transformation is also able to carry through the spatial modes contained in the observed field as shown in Figure \ref{fig:EOF_0}. The efficacy of the transformation is stronger for the HYCOM than for the MITgcm-GoM field, which also shows that there is room for improvement in the data assimilation process, whether in the NCODA system or by applying our Transform Model.

\subsubsection {Three-dimensional field transformation}
Surface observations are in general more commonly available than subsurface observations, in particular from three-dimensional arrays. Therefore, predicting subsurface dynamics has remained more challenging than at the ocean surface because of the lack of continuous sampling at the same location. Vertical projections of the surface corrections is still reliant on methods described in \cite{Cooper1996AltimetricAW}, \cite{Fox2002}. The transformation of three-dimensional fields is therefore critical to full water column prediction. For the three-dimensional velocity field transformation, a Transform Model was created for each depth level between the numerical model and the observations and the transformation was applied to the last 50 days not included in the training. 

\begin{figure}[ht]
\begin{subfigure}{.95\textwidth}
  \centering
  \includegraphics[trim={0 0cm 0 0cm},clip,width=.95\linewidth]{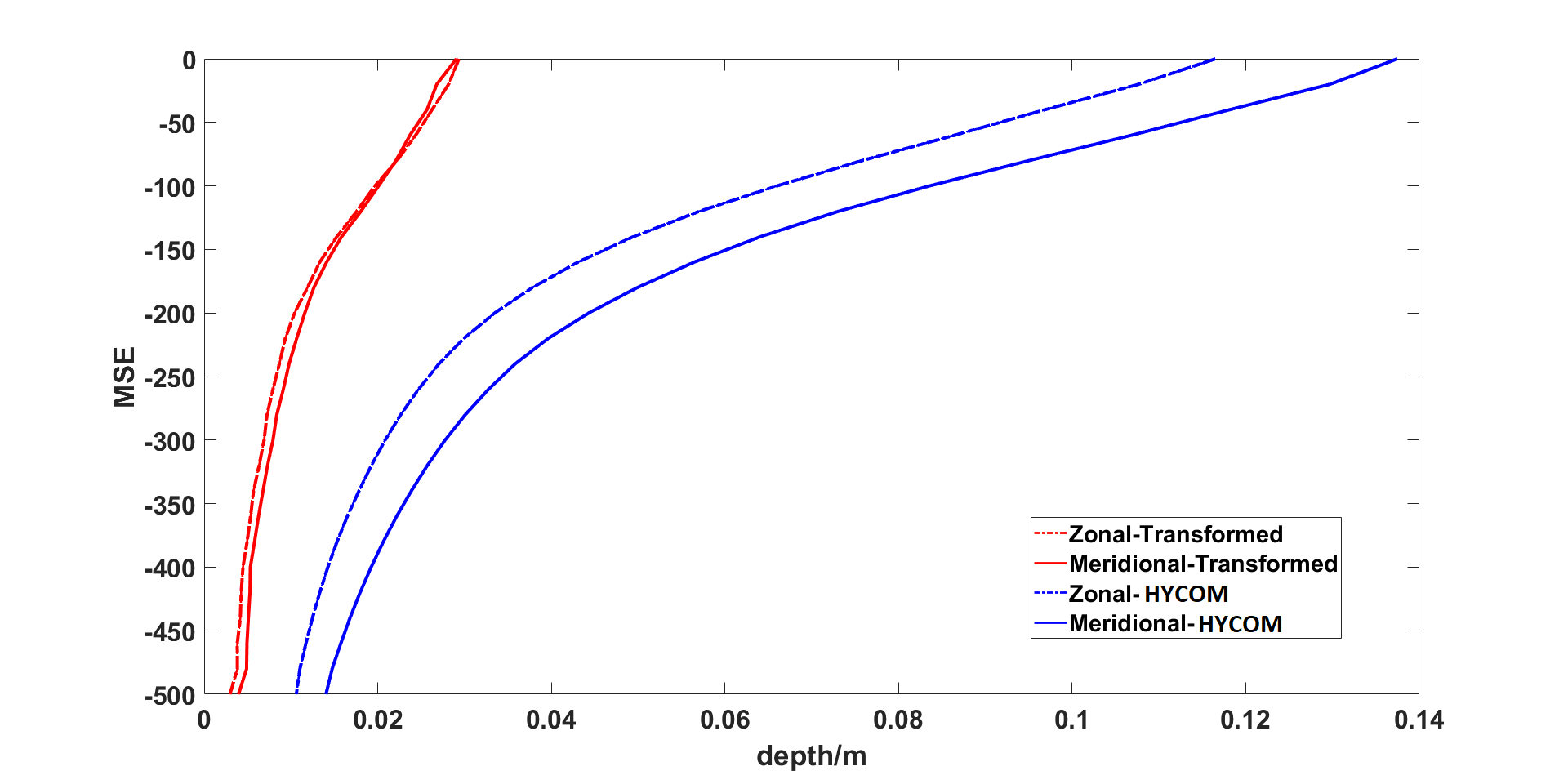}
  \caption{Subsurface MSE }
  \label{fig:HYCOMdepth_RMSE}
\end{subfigure}
\begin{subfigure}{.95\textwidth}
  \centering
 \includegraphics[trim={0 0cm 0 0cm},clip,width=.95\linewidth]{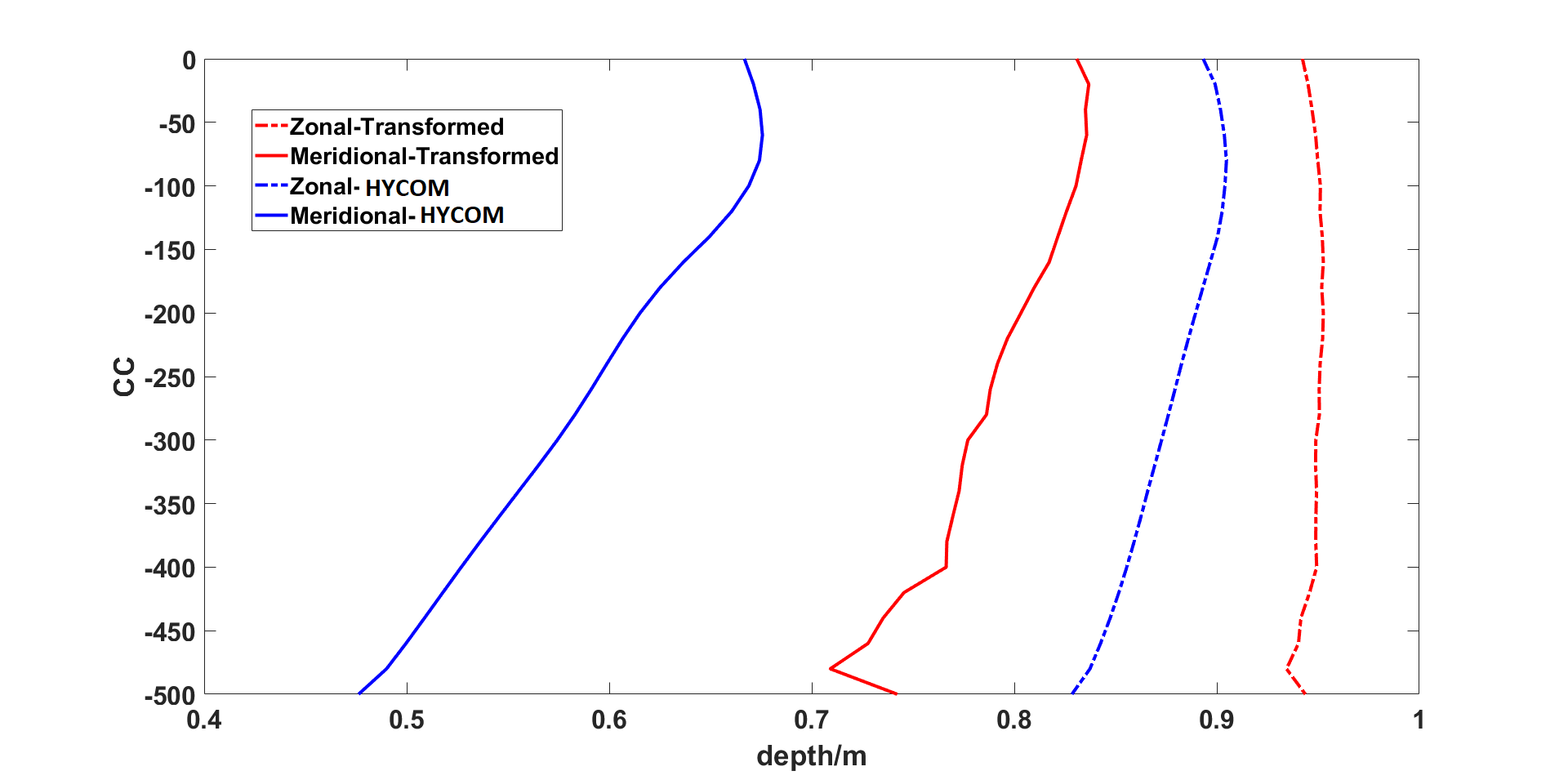}\\ 
  \caption{Subsurface CC}
  \label{fig:HYCOMdepth_CC}
\end{subfigure}
\caption{Subsurface velocity average mean squared error (MSE - a) and correlation coefficient (CC - b) between the original and the transformed HYCOM velocity tensor time series.}
\label{fig:HYCOM_RMSE_CC}
\end{figure}

The results in terms of MSE and correlation coefficient (CC) for each depth level for the transformed HYCOM model over the 50-day period are shown in Figure \ref{fig:HYCOM_RMSE_CC}. As seen in the two-dimensional case, the transformation efficacy is significant. The MSE is reduced by a factor four from the surface down to 200m and by half at the 500m level (Figure \ref{fig:HYCOMdepth_RMSE}). The HYCOM field's MSE decreases with depth along with the gain of the transformation (Table \ref{tab:correction}). While the MSE of the HYCOM meridional velocity field is higher than the one of the zonal velocity field, when transformed, the resulting MSEs of each components are nearly identical. Similarly, the CC of the meridional field is much less than the CC of the zonal field across all levels (Figure \ref{fig:HYCOMdepth_CC}). The difference persists in the transformed fields but is strongly reduced. In addition the CC of the transformed fields is larger than the CC of the original fields and varies less vertically as well. The difference between zonal and meridional velocity CC is not surprising since the current meridional velocities dominate the flow pattern of the Loop Current.

\begin{figure}[ht]
\begin{subfigure}{.95\textwidth}
  \centering
  \includegraphics[trim={0 0cm 0 0cm},clip,width=.95\linewidth]{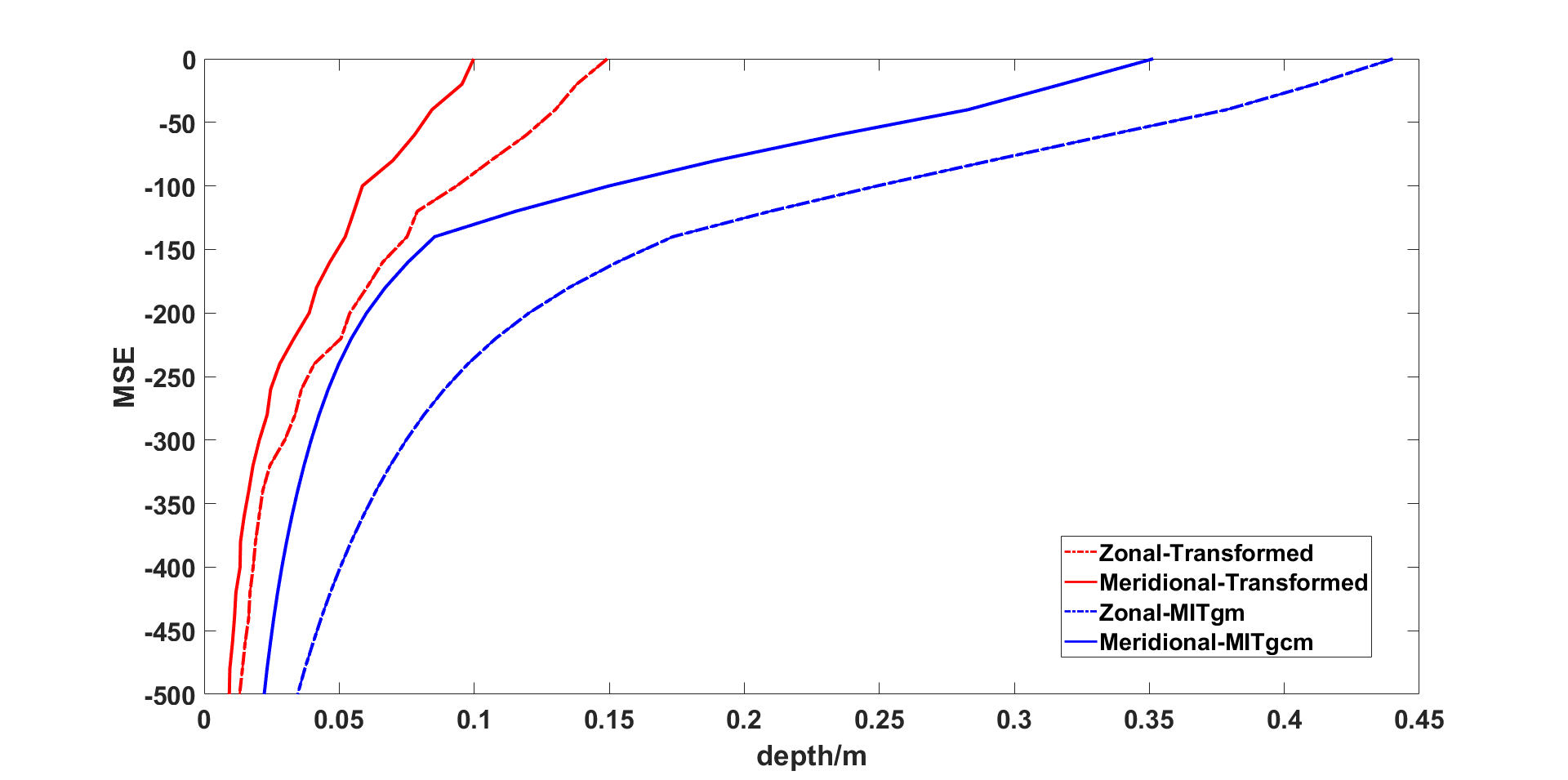}
  \caption{Subsurface MSE }
  \label{fig:Depth_RMSE}
\end{subfigure}
\begin{subfigure}{.95\textwidth}
  \centering
 \includegraphics[trim={0 0cm 0 0cm},clip,width=.95\linewidth]{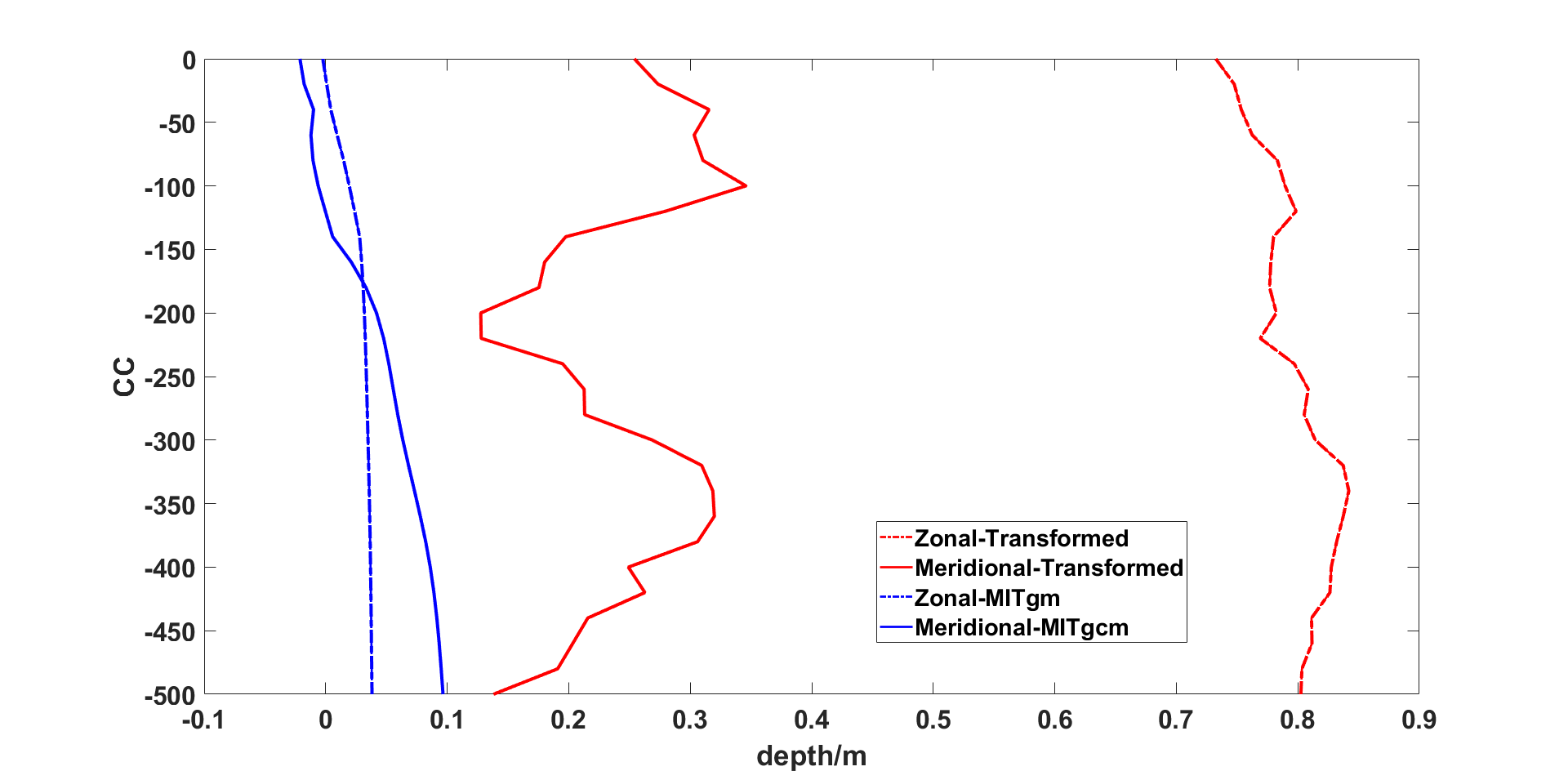}
  \caption{Subsurface CC}
  \label{fig:Depth_CC}
\end{subfigure}
\caption{ Same as Figure \ref {fig:HYCOM_RMSE_CC} but for MITgcm-GoM  model.}
\label{fig:ECCO_RMSE_CC}
\end{figure}

For the MITgcm-GoM three-dimensional transformed velocity fields the MSE is reduced by a factor three from the surface down to 200m and by half at the 500m level (Figure \ref{fig:Depth_RMSE}) although that reduction is less for the meridional velocity below 150m. For the CC, the MITgcm-GoM original model show very small correlation with the observations. The transformed fields however, shows an increase for the zonal velocity by a factor sixteen (Figure \ref{fig:Depth_CC}), relatively constant over depth, in contrast with the CC of the transformed meridional velocity, which is not as uniformely improved, although the correction can reach a factor fifteen near 100 and 350m. The 50-days gain is therefore less for the MITgcm-GoM overall as shown by Table \ref{tab:correction} but the correction factor is much higher for the latter than for HYCOM. The MSE of the transformed MITgcm-GoM fields (red lines Figure \ref{fig:Depth_RMSE}) is similar to the HYCOM's original MSE (blue lines in Figure \ref{fig:HYCOMdepth_RMSE}). From a correlation standpoint, there is a similar improvement, in particular for the zonal velocity (red lines in Figure \ref{fig:Depth_CC} vs blue lines in Figure \ref{fig:HYCOMdepth_CC}) and to a lesser degree for the meridional velocity. This shows that our method is at least as efficient at correcting the free-run model with just the observed velocity field as the NCODA is with all the different data sources used for HYCOM. The latter on the other hand requires less correction by the Transform Model, but the improvements by the Transform Model can still be significant, especially near the surface.

\begin{table}[htbp]
\caption{The average percentile gain of the velocity corrections at different depths by applying the Transform Model to two data time series at the 50th day of the correction process.}
\begin{center}
\begin{tabular}{|c|c|c|c||c|c|c|} \hline
\textbf{Dataset}&{\textbf{Zonal}}&{\textbf{Zonal}}&{\textbf{Zonal}}&{\textbf{Meridional}}&{\textbf{Meridional}}&{\textbf{Meridional}}\\ \hline
\textbf{Depth}&{\textbf{0 m}}&{\textbf{-100m}}&{\textbf{-500m}} &{ \textbf{0 m}}&{\textbf{-100m}}&{\textbf{-500m}}\\ \hline

\textbf{HYCOM} &$ 78.85 \%$ &$76.23 \%$ &$  71.78 \%$ &$74.80 \%$ &$ 69.16 \%$ &$ 71.99\% $  \\\hline
\textbf{MITgcm}  &$54.88 \%$ &$ 43.43 \%  $ &$ 35.53 \% $ &$48.35 \% $ &$ 21.29 \% $ &$ 42.91\% $  \\\hline
\end{tabular}
\label{tab:correction}
\end{center}
\end{table}

\begin{figure}[ht]
\begin{subfigure}{.95\textwidth}
  \centering
   \includegraphics[trim={0 0cm 0 0cm},clip,width=.9\linewidth]{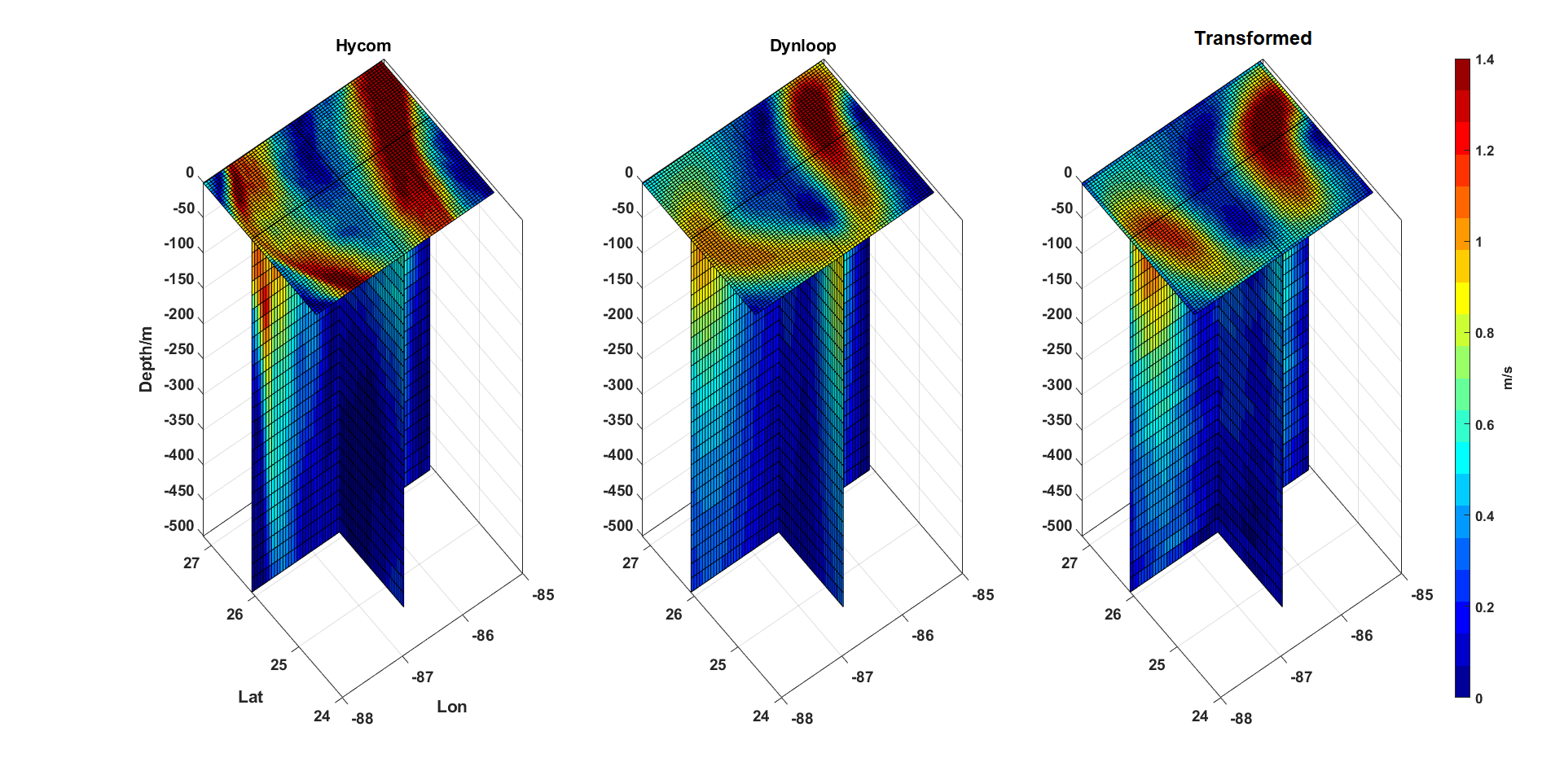}
  \caption{}
  \label{fig:3d100}
\end{subfigure}
\begin{subfigure}{.95\textwidth}
  \centering
 \includegraphics[trim={0 0cm 0 0cm},clip,width=.9\linewidth]{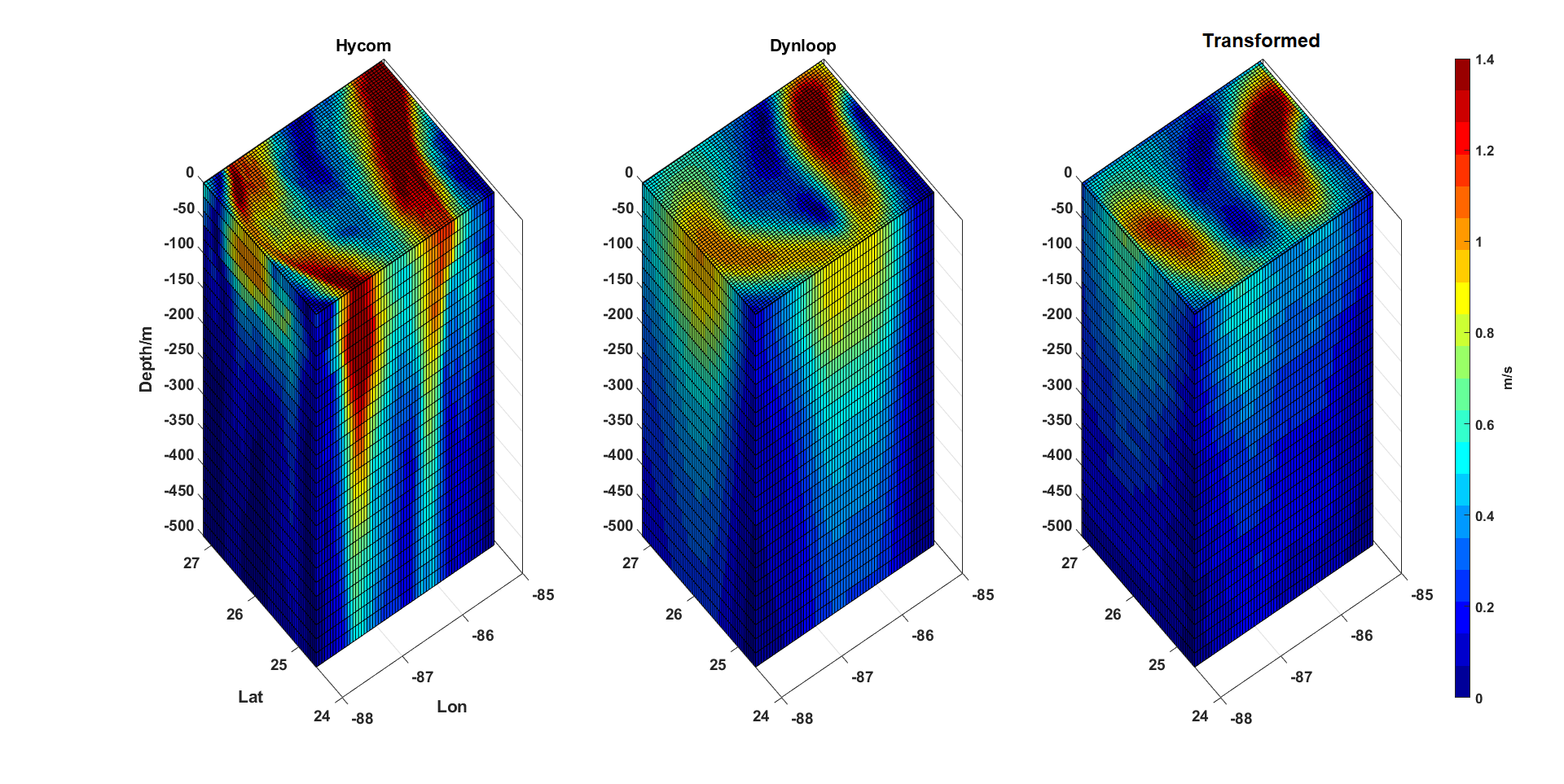}\\ 
  \caption{}
  \label{fig:3d200}
\end{subfigure}
\caption{Transformed HYCOM three-dimensional velocity structures on day 50 after the end of the observation period. Both the structures inside (a) and on the boundary (b) of the transformed volume are shown. The left, middle and right plots show the original HYCOM, the Dynloop, and the transformed flow fields.}
\label{fig:3DHYCOM_trans}
\end{figure}

\begin{figure}[ht]
\begin{subfigure}{.95\textwidth}
  \centering
  \includegraphics[trim={0 0cm 0 0cm},clip,width=.9\linewidth]{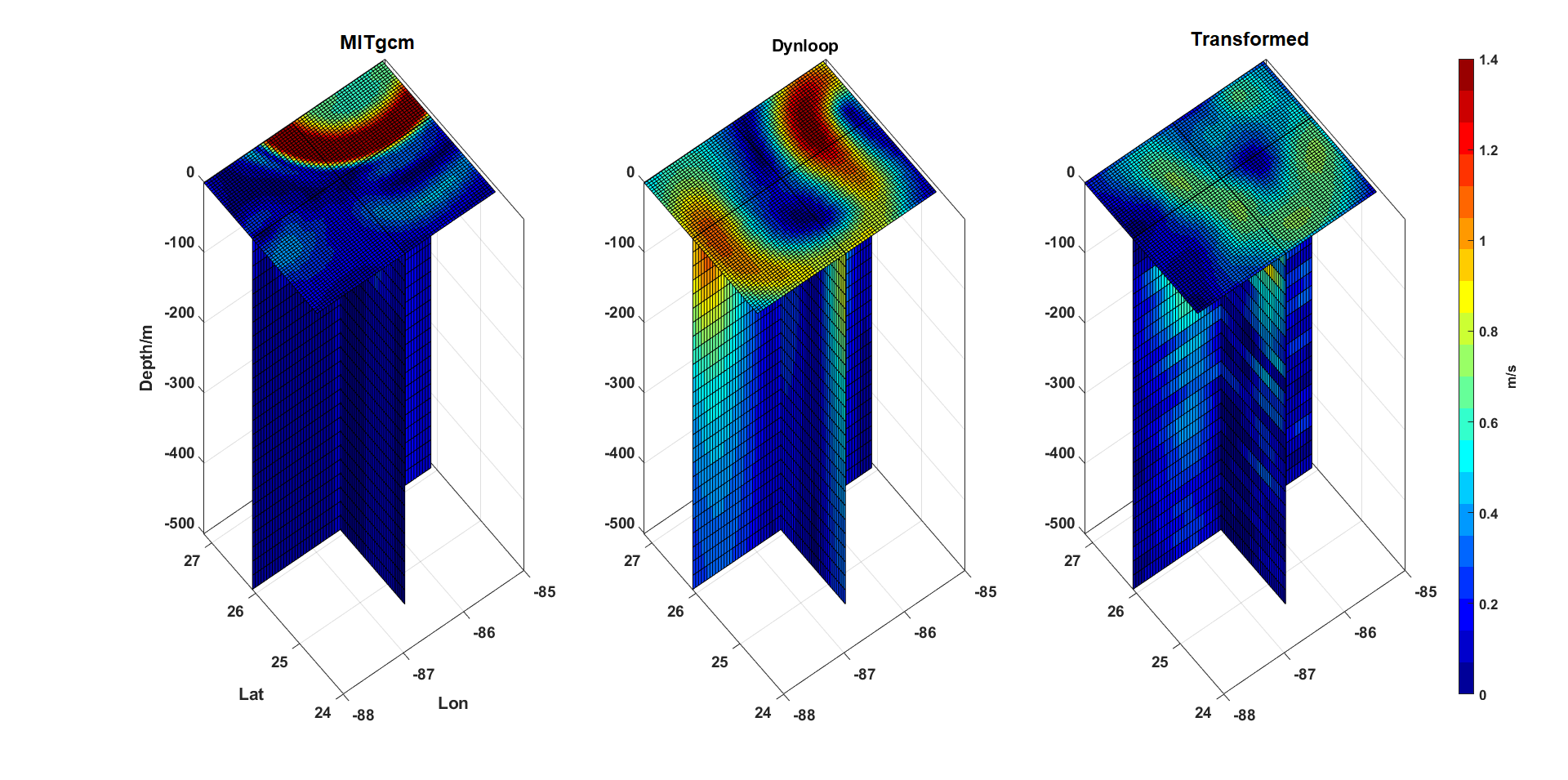}
  \caption{}
  \label{fig:Day_50}
\end{subfigure}
\begin{subfigure}{.95\textwidth}
  \centering
 \includegraphics[trim={0 0cm 0 0cm},clip,width=.9\linewidth]{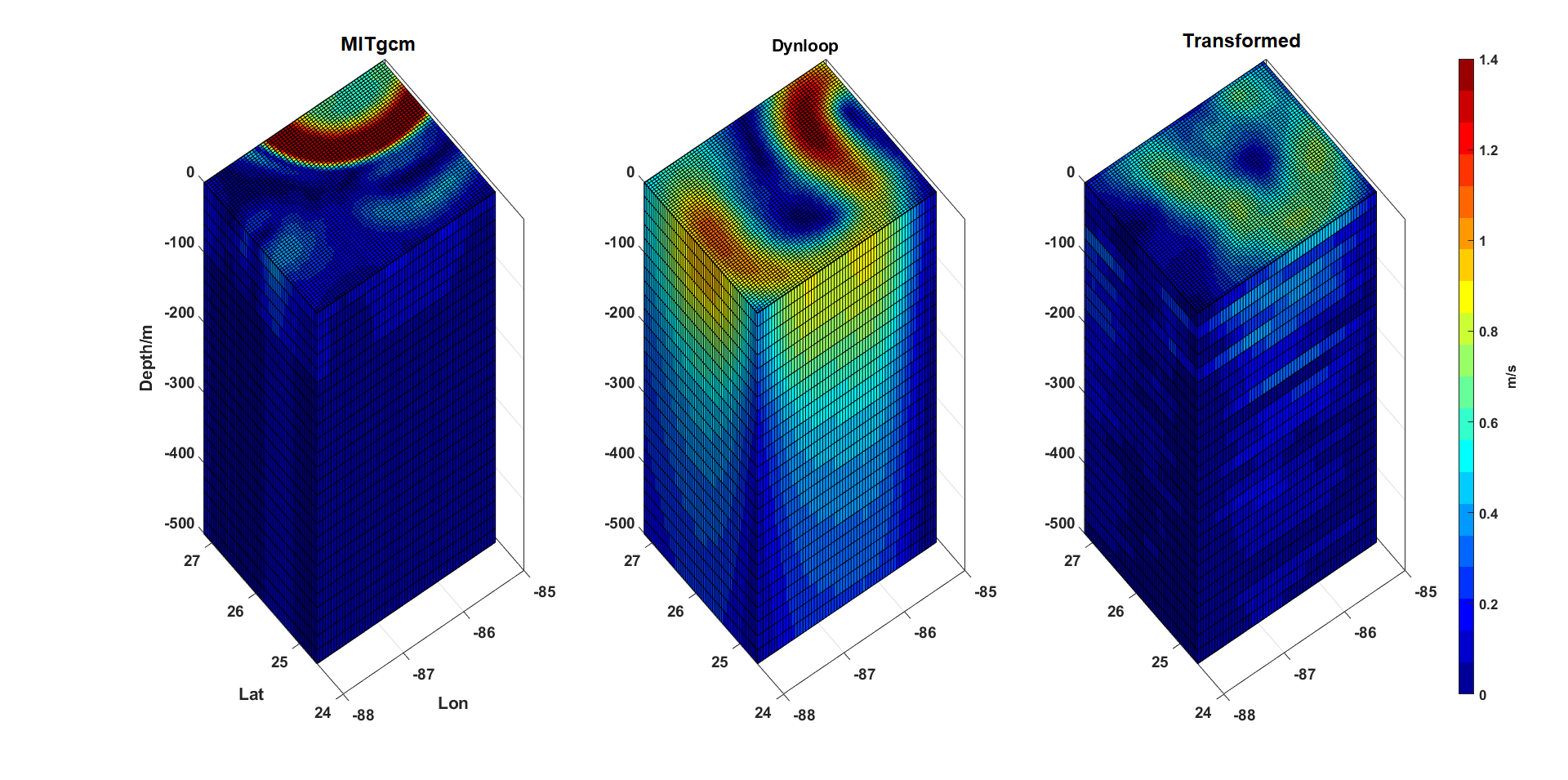}\\ 
  \caption{}
  \label{fig:ThreeD_50}
\end{subfigure}
\caption{ Same as Figure  \ref{fig:3DHYCOM_trans} but for MITgcm-GoM model}
\label{fig:ThreeD_ECCO}
\end{figure}

While the statistical metrics such as MSE and CC may indicate a good agreement between the quantities compared, the spatial features that comprise the system may look dissimilar. Therefore, we show here the efficacy of the transformation in the three-dimensional structures of the transformed field on day 50 after the end of the observation period (Figure \ref{fig:3DHYCOM_trans}, \ref{fig:ThreeD_ECCO}). For the HYCOM model, the vertical and horizontal flow features in the observations are present in the transformed model field, showing similarities in magnitude and location in space. The presence of these features and their timing is critical to the development of the instabilities that lead to the growth of the frontal cyclones near the surface and then in the deep waters and ultimately to the separation of a Loop Current eddy \cite{donohue2016a}, \cite{DONOHUE2016b}, \cite{cherubin06}, \cite{cherubin05}. The three-dimensional transformation confirms the results seen in the two-dimensional transformation. The Transform Model is capable of extending the correction of observations in HYCOM at least 50 days past their the last measurement. For the MITgcm-GoM model, flow field difference is quite significant as shown by Figure \ref{fig:ThreeD_ECCO}. The transformed flow field is shifted toward the path of the observed flow, both horizontally and vertically, although the magnitude of the transformed flow is reduced from the observations.




\subsection{Long-term Transform Model Evaluation} 
\begin{figure}[ht]
\begin{subfigure}{.95\textwidth}
  \centering
  \includegraphics[trim={0 0cm 0 0cm},clip,width=.95\linewidth]{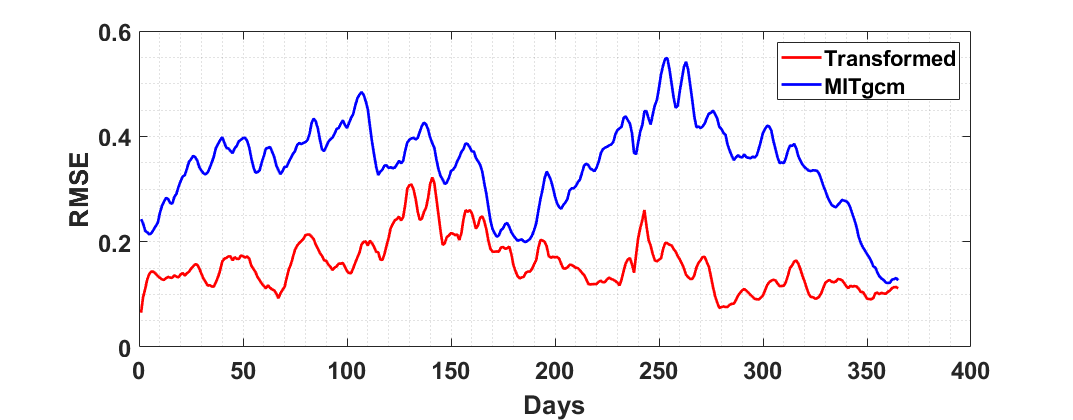}
  \caption{}
  \label{fig:Future_RMSEM}
\end{subfigure}
\begin{subfigure}{.95\textwidth}
  \centering
 \includegraphics[trim={0 0cm 0 0cm},clip,width=.95\linewidth]{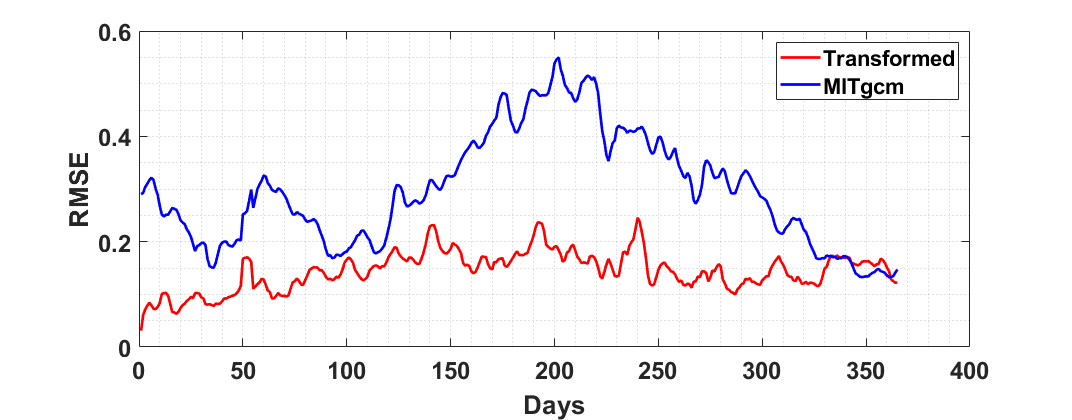}\\ 
  \caption{}
  \label{fig:Past_RMSEM}
\end{subfigure}
\caption{One year surface velocity time series MSE for the original (blue line) and transformed (red line) MITgcm-GoM in the (a)forward direction  and (b) backward direction.}
\label{fig:ECCO_RMSE_F_P}
\end{figure}

For this evaluation HYCOM is used as virtual observations to transform the MITgcm-GoM surface velocity field. In the forward transformation experiment the transformation was applied for up to 365 days after the virtual observations ended. As seen in the short-term transformation, the MSE of the transformed field is reduced by up to a factor four and exhibits less temporal variability than the MITgcm-GoM MSE (Figure \ref{fig:ECCO_RMSE_F_P}). The MSE of the transformed field varies over time along with the evolution of the Loop Current System, whose dynamics can affect the efficacy of the transformation. Indeed between days 120 and 160, the Loop Current evolution between the HYCOM and the MITgcm-GoM significantly diverged. While the eddy shed and reattached in the HYCOM flow, the Loop Current remained in its extended position in the MITgcm-GoM model. This event affected the capacity of the Transform Model to correct the divergent dynamics, which increased the MSE of the transformed model. Nonetheless, the transformation is able to strongly reduce the MITcgm-GoM velocity field MSE even 300 days after the virtual observation ended, which reveals the long lasting potential of the transformation process. The transformation is most efficient when the difference between models is associated with a spatial shift of the flow features (days 200-365 on Figure \ref{fig:Future_RMSEM}) and not a dynamical state difference. 

The backward transformation also shows a strong decrease in the MSE of the transformed MITgcm-GoM velocity fields for most part except during periods when the original numerical model MSE is also low. In this case, both MITgcm-GoM and HYCOM circulation features exhibit dynamical similarities. In addition the transformed MSE is less variable than the original one. 

\begin{figure}[ht]
\begin{subfigure}{.95\textwidth}
  \centering
  \includegraphics[trim={0 0cm 0 0cm},clip,width=.85\linewidth]{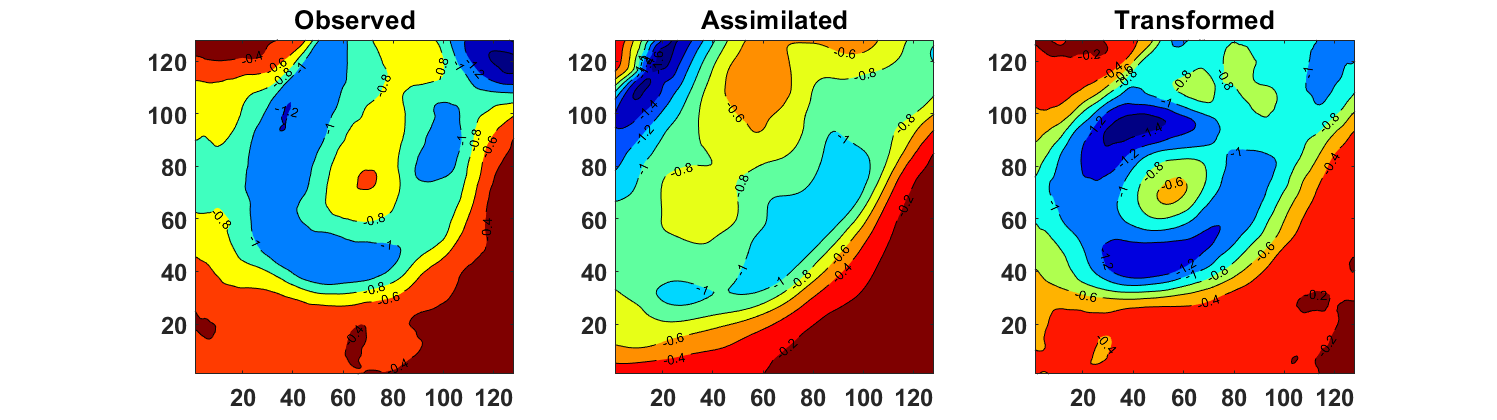}
  \caption{}
  \label{fig:Future_EOF}
\end{subfigure}
\begin{subfigure}{.95\textwidth}
  \centering
 \includegraphics[trim={0 0cm 0 0cm},clip,width=.85\linewidth]{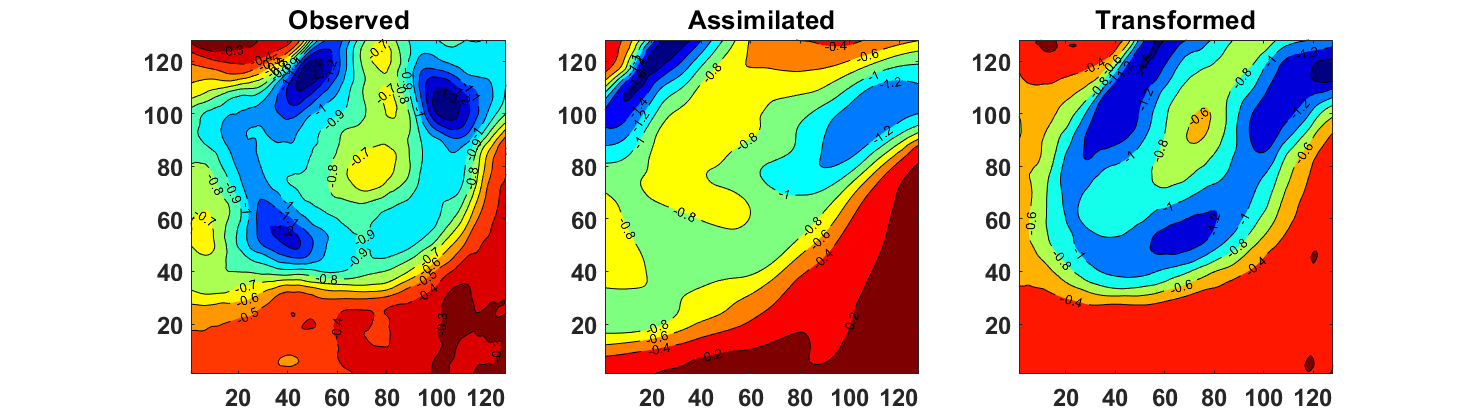}\\ 
  \caption{}
  \label{fig:Past_EOF}
\end{subfigure}
\caption{First Empirical Orthogonal Function (EOF) mode of the surface velocity fields over the first (last) 120-day of transformed field in the forward (a) and backward (b) direction. From left to right is shown the observed, the model original, and the transformed field.}
\label{fig:ECCO_EOF_F_P}
\end{figure}

\begin{figure}[ht]
\centering
\includegraphics[trim={0 0cm 0 0cm},clip,width=.9\linewidth]{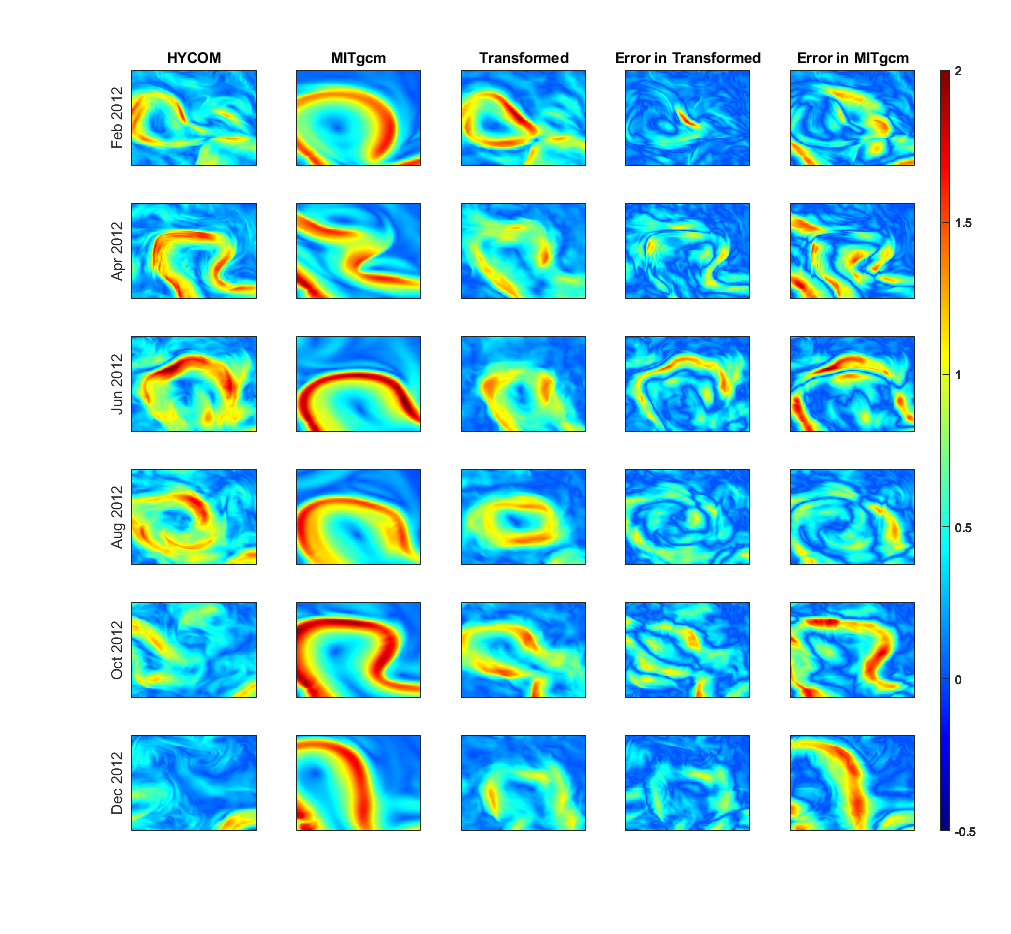}
\caption{Forward Transform Model - Bi-monthly (rows from February to December 2012) velocity magnitude of the flow field of the virtual observation (HYCOM - first column on the left)), of the MITgcm-GoM (second column), and of the transformed MITgcm-GoM (third column), respectively. The last two columns show the error field of the transformed model (fourth column) and of the MITgcm-GoM model, respectively.}
\label{Future_Samples}
\end{figure}

\begin{figure}[h!]
\centering
\includegraphics[trim={0 0cm 0 0cm},clip,width=.9\linewidth]{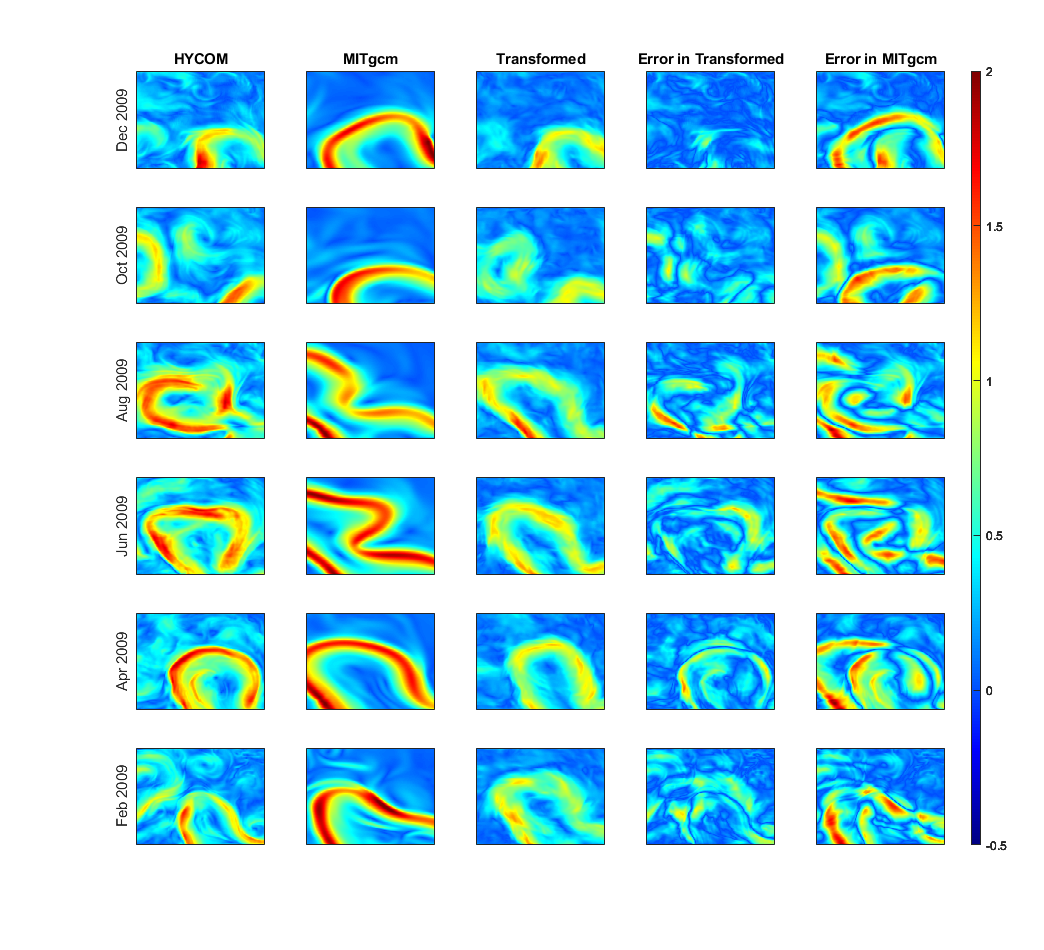}
\caption{Same as Fig. \ref{Future_Samples} but for the backward Transform Model.}
\label{Past_Samples}
\end{figure}

Although the transformation could lead to unrealistic features after several months, the EOF analysis of the velocity field reveals that the modal structure of the virtual observations is present in the transformed field, both in the forward and backward transformations (Figure \ref{fig:ECCO_EOF_F_P}). The transformation is also able to convey the timeliness of the separation of the Loop Current eddy (Figure \ref{Future_Samples}) in the forward transformation, which is paramount to the generation of reliable and useful predictions of the Loop Current. In the backward direction, the Transform Model is capable of readjusting the Loop Current state in the MITgcm-GoM to closely follow the state of the Loop Current in the virtual observations (Figure \ref{Past_Samples}).

\section*{Conclusions}\label{sec:Conclusion}

Reliable and long-term ocean observations are difficult to obtain due to a variety of factors such as but not limited to costs, data collection manned or unmanned platform availability, instruments failure and durability, accessibility, limited energy supplies, data storage and extraction. In the current state of environmental predictions, data used for prognostic activities such as ocean or weather forecast are only used at the time of the forecast. They become part of history as soon as the forecast is done and are no longer relevant to the forecast. What if we could extend the period of relevance both forward and backward in time, in other words the usability of environmental observations? In this study we have presented a new method based on deep learning that enables the extension of the usability period of real-world observations up to one year. We have focused on ocean observations, but we believe that our approach could be applied to any biogeochemical and physical prediction exercise, as long as observations and model output are concurrently available.

Our approach is based on the re-purpose of an existing deep learning model, called U-Net, designed specifically for image segmentation analysis in the biomedical field. We have modified the network in such a way that its original segmentation layer is replaced by, instead a regression layer. That layer is used to create a Transform Model that retains the temporal and spatial evolution of the differences between model and observations to produce a correction in the form of regression weights that evolves spatially and temporally with the model both forward and backward in time, beyond the observation period. In classical data-assimilation problems, the corrections that the observations make to the first-guess/background must lie in the space spanned by $P$, the error covariance matrix. As such, $P$ has been the subject of much research because of the central role it plays in data assimilation \cite{Moore2019}. For example, estimating the actual level of uncertainty of the first-guess is very difficult, and choosing a $P$ that accurately reflects the inhomogeneity and anisotropic nature of the errors across the broad range of space- and time-scales that characterize the ocean is challenging, although innovative methods for multi-scale data-assimilation are being explored \cite{Li2015}, \cite{Mirouze2016}. In addition there is no efficient method to spread information forward or backward in time and to evolve the error covariance matrix in time, which can be attempted by using ensemble-based Khalman filters methods \cite{EnKF2016}. Even though we haven't applied our approach to assimilate ocean data in numerical models, our method could be applied to the estimation of $P$ while solving the time dependency, the spatial inhomogeneity and anisotropy and at the same time increase the usability of observations beyond their time measurement for up to maybe one year as shown in this study. It would also be applicable to the many other challenges faced by data-assimilation, associated with model and observations errors. The evolving relationship between the best model estimate and the main causes of errors including model and instrument errors could be captured by a Transform Model as described in this study without previous knowledge of absent physics or small scale processes not resolved by the model. 

Ultimately, one could envision that the first-guess estimate could be entirely realized with our Transform Model. Our transformation could be applied sequentially, at each time step of the prediction, which would reduce the magnitude of the increments and the shock induced to the numerical model \cite{RAGHUKUMAR2015546}, by keeping the model on a track closer to the observations than when no correction has been applied to the previous step as done in this work. While very few data-assimilation  systems consistently assimilate sub-surface ocean current velocity data, the results shown in this study also reveal the efficacy of the Transform Model at correcting the three-dimensional flow field, even on existing reanalyzes, such as the HYCOM data used herein. Finally, with the increased availability of surface current data from high-frequency (HF) radars, a Transform Model could be easily created for any numerical model with just a few years of concurrent model data and observations. Such model, which can be updated daily with new data would provide useful corrections to the model surface flow field that would be transferred vertically to other variables, either by the numerical model itself or through a similar type of AI technique as the one described here.

\bibliography{scifile}

\bibliographystyle{Science}


\section*{Supplementary materials}
Figs. S1 to S14\\

\end{document}